\documentclass[10pt,%twocolumn,
prl,aps,showpacs]{revtex4}
\usepackage{amsmath}
\usepackage{amssymb}
\usepackage{amsfonts}
\usepackage{graphicx}
\usepackage{dcolumn}
\usepackage{hyperref}
\newcommand{\be}{\begin{equation}}
\newcommand{\ee}{\end{equation}}
\newcommand{\bea}{\begin{eqnarray}}
\newcommand{\eea}{\end{eqnarray}}
\newcommand{\beaa}{\begin{eqnarray*}}
\newcommand{\eeaa}{\end{eqnarray*}}

\newcommand{\e}{\mathrm{e}}
\begin{document}
\title{Scalar-tensor representation of $f(R)$ gravity and Birkhoff's theorem}
\author{S. Capozziello$^{1,2}$ and D. S\'aez-G\'omez$^{3}$}
\affiliation{$^{1}$Dipartimento di Scienze Fisiche, Universit\`a di Napoli ``Federico II'' and \\
$^{2}$INFN Sez. di Napoli, Compl. Univ. Monte S. Angelo, Ed.N, Via Cinthia, I-80126 Napoli, Italy, EU}
\affiliation{$^{3}$Institut de Ci\`encies de l'Espai (ICE-CSIC/IEEC),
Campus UAB, Facultat de Ciencies, Torre C5-Par-2a pl, E-08193
Bellaterra (Barcelona), Spain, EU}

\begin{abstract}
Birkhoff's theorem is discussed in the frame of $f(R)$ gravity by using its scalar-tensor representation. Modified gravity has become very popular at recent times as it is able to reproduce the unification of inflation and late-time acceleration with no need of a dark energy component or an inflaton field. Here, another aspect of modified $f(R)$ gravity is studied, specifically the range of validity of Birkhoff's theorem,  compared with another alternative to General Relativity, the well known Brans-Dicke theory. As a novelty, here  both theories are studied  by using a conformal transformation and writing the actions in the Einstein frame, where spherically symmetric solutions are studied by using  perturbation techniques. The differences between both theories are analyzed as well as the validity of the theorem within the Jordan and Einstein frames, where interesting results are obtained.
\end{abstract}

\maketitle

\section{Introduction}

In the last decade, modified theories of gravity have become very popular as they can explain the accelerating expansion of the  Universe  without  need to introduce dark energy. Such theories can even provide an alternative description of inflation without any need for an inflaton field. Hence,   dark energy and inflationary eras appear as a natural consequence of the  law of gravity, which has a different behavior depending on the scale. Then,  gravity is supposed to be well described by General Relativity at local scales, while  the additional terms introduced in theories of modified gravity become important at cosmological scales, so that the cosmological evolution can be described in  purely gravitational terms with no need to assume additional components (a seminal paper in this sense is \cite{curvature}). Special interest in this gravitational paradigm  is related with $f(R)$ gravity (for  general reviews, see \cite{review,book}), which has a simple structure, where more complex functions of the Ricci scalar than in the Hilbert-Einstein action are considered, but with no other curvature invariant or non-local term. Within $f(R)$ gravity, reconstruction of inflationary and dark energy epochs can be easily performed and useful techniques have been developed in order to reconstruct the appropriate action that can describe the Universe evolution (see \cite{reconstruction1}). Also, the so-called viable models of $f(R)$ gravity can avoid violations of local gravity tests and matter instabilities (see Ref.~\cite{f(R)viable1}). In this sense, the study of spherically symmetric solutions in the frame of $f(R)$ gravity turns out an important and interesting question, which has been already dealt in Ref.~\cite{SphericalFR}, as well as the study of the Newtonian limit of the theory, which can constrain the form of the action (see Ref.~\cite{Capozziello3}). In most cases the reconstruction of the action $f(R)$, specially when dealing with Friedmann-Lema\^itre-Robertson-Walker metrics, is done by using  an auxiliary scalar field, since $f(R)$ gravity is equivalent to a kind of Brans-Dicke theory with a non-propagating scalar field and a non-null potential. However, both theories seem to exhibit a different behavior in the study of perturbations, as it was pointed out in Ref.~\cite{Capozziello1}. This is probably due to the high non-linearity exhibited in $f(R)$ gravities, which can be alleviated when one deals with its equivalent picture in a Brans-Dicke-like theory.   \\

At the present paper, we are interested to study spherically symmetric solutions and specifically Birkhoff's theorem in $f(R)$ gravity and its scalar-tensor representation. It is well known that,  in $f(R)$ gravity as in Brans-Dicke theory, Birkhoff's theorem is not satisfied unless strong restrictions are imposed on the scalar curvature and on the scalar field respectively, Ref.~\cite{Faraoni}. However, here we are interested to study the range of validity of Birkhoff's theorem  by performing perturbations around a background solution in $f(R)$ gravity, where an auxiliary scalar field is used, and compare the results with those for  Brans-Dicke theory. In order to obtain a real comparative of the perturbations  in both theories, we shall use the mathematically equivalent picture defined in the Einstein frame, which is related with the original one, usually called Jordan frame, by a conformal transformation. Then, by writing $f(R)$ gravity and Brans-Dicke theory in the Einstein frame, we study the perturbations around a background solution, which can be seen as a new technique to compare both theories. The results obtained in the Einstein and the Jordan frames are studied, where it is found different information on the Birkhoff's theorem for each frame, what may suggest the non-physically equivalence between both, as it was already pointed out in Ref.~\cite{JordanVsEinstein} in a cosmological context.\\

Hence, the paper is organized as follows: next section is aimed to the framework of the paper, where $f(R)$ gravity and its scalar-tensor representation (the so-called O'Halon theory) and Brans-Dicke theory are introduced. The relation between the Jordan frame, where both theories are defined, and the Einstein frame is briefly introduced in Section III, where the transformation of a general spherically symmetric solution is obtained. In  section IV, the perturbative study of Birkhoff's theorem in the Einstein frame  is done, where the main results of the paper are obtained. Finally in section V, we summarize and present some conclusions.

\section{The framework}
Let us start  writing the action and field equations that describe a general $f(R)$ theory. We consider the action,
\be
\label{1.1}
S= \int d^4x \sqrt{-g} \left[f(R) + 2\kappa^2\mathcal{L}_\mathrm{m} \right]\ .
\ee
where $\kappa^2=8\pi G$, $R$ is the Ricci scalar and $\mathcal{L}_\mathrm{m}$ stands for the Lagrangian corresponding to matter of some kind. Note that this action reduces to General Relativity for $f(R)=R$, and  the corresponding field equations turn out of second order, while for the more general action (\ref{1.1}), the equations are fourth order, which are obtained  by varying the action (\ref{1.1}) respect to the metric tensor $g_{\mu\nu}$,
\be
  R_{\mu\nu} f'(R)- \frac{1}{2} g_{\mu\nu} f(R) + g_{\mu\nu}  \Box f'(R) -  \nabla_{\mu} \nabla_{\nu}f'(R)=\kappa^2T_{\mu\nu}\ .
\label{1.2}
\ee
Here the primes denote derivatives with respect $R$, and the energy-momentum tensor is given by $T_{\mu\nu}^{(m)}=\frac{-2}{\sqrt{-g}}\frac{\delta\mathcal{L}_\mathrm{m}}{\delta g^{\mu\nu}}$. By taking the trace of the equation (\ref{1.2}), one obtains an extra equation given by,
\be
3\Box f'(R)+f'(R)R-2f=\kappa^2T_{\mu\nu}^{(m)}\ .
\label{1.3}
\ee
Due to the extra terms in the field equations, these theories can reproduce quite well the cosmic acceleration with no need to introduce a dark energy component or a cosmological constant, and even the inflationary epoch can be described as a pure gravitational effect (for some literature, see Refs.~\cite{reconstruction1} and \cite{f(R)viable1}). The study of this class of theories in contexts apart from cosmology is a fundamental issue to understand these theories and to provide new results. At the current paper we are interested to study spherically symmetric solutions (see Ref.~\cite{SphericalFR}-\cite{Capozziello1}) and the range of validity of the Birkhoff's theorem in $f(R)$ gravity. We will explore the relation between $f(R)$ gravity  and scalar-tensor theory in the Jordan and Einstein frames, and we compare the results obtained for each frame. For that, let us write the general action for a Brans-Dicke-like theory,
\be
S_{BD}=\int d^4x\sqrt{-g}\left[\phi R-\frac{w}{\phi}g^{\mu\nu}\nabla_{\mu}\phi\nabla_{\nu}\phi-V(\phi)+2\kappa^2\mathcal{L}_m\right]\ .
\label{1.4}
\ee
Here we assume $w$ to be a constant. This action is written in the so-called Jordan frame, which is related with the Einstein frame by means of a conformal transformation. The field equations for this action are obtained by varying the action respect the metric tensor  $g_{\mu\nu}$ and the scalar field $\phi$,
\[
R_{\mu\nu}-\frac{1}{2}g_{\mu\nu}R=\frac{\kappa^2}{\phi}T_{\mu\nu}^{(m)}+\frac{w}{\phi^2}\left[\nabla_{\mu}\phi\nabla_{\nu}\phi-\frac{1}{2}g_{\mu\nu}\nabla^{\sigma}\phi\nabla_{\sigma}\phi\right]+\frac{1}{\phi}(\nabla_{\mu}\phi\nabla_{\nu}\phi-\Box\phi)-g_{\mu\nu}V(\phi)\ ,
\]
\be
(2w+3)\Box\phi=\kappa^2T^{(m)}+\phi\frac{dV(\phi)}{d\phi}-2V(\phi)\ .
\label{1.5}
\ee
The action (\ref{1.4}) becomes the Lagrangian describing O'Halon theory for a null kinetic term $w=0$, such that the actions (\ref{1.1}) and (\ref{1.4}) turn out equivalent. This is straightforward to see, just by varying the action (\ref{1.4}) with respect the scalar field $\phi$, where the relations $f'(R)=\phi$ and $V'(\phi)=R$ are obtained, which reflects the extra scalar degree of freedom that $f(R)$ gravity owns, where the trace equation (\ref{1.3}) can be seen as a Klein-Gordon equation. On the other hand, the original Brans-Dicke theory is given by taking a null scalar potential $V(\phi)=0$ and  $w\neq0$ in the action (\ref{1.4}). Despite the similarity between the two theories, when the Newtonian limit is performed,  the well known PPN parameter for Brans-Dicke theory, given by $\gamma_{BD}=\frac{1+w}{2+w}$, does not correspond to the PPN parameter for the O'Halon theory (equivalent to $f(R)$ gravity) by taking $w=0$, which gives $\gamma=1/2$, but the PPN parameter for $f(R)$ gravity is given by a more complex expression  as it is shown in Ref.~\cite{Capozziello1}. Nevertheless, here we are interested to study perturbations around a given background solution, particularly we will study spherically symmetric solutions of the type,
\be
ds^2=-A(r,t)dt^2+B(r,t)dr^2+r^2d\Omega\ ,
\label{1.6}
\ee
where $d\Omega$ is the metric of a 2-sphere. The metric (\ref{1.6}) can be considered as the most general spherically symmetric solution, where the coordinates have been chosen to avoid crossed terms between the spatial and time terms. For the theories described by the field equations (\ref{1.5}), the only way to ensure a static  solution (\ref{1.6}), that is $A(r,t)=A(r)$ and $B(r,t)=B(r)$, is to impose a time independent  scalar field $\phi(r,t)=\phi(r)$, which implies in the equivalent $f(R)$ gravity representation to assume a time independent Ricci scalar in the field equations (\ref{1.2}) (see Ref.~\cite{Faraoni}). At the present paper, we do not impose any condition on the scalar field to study spherically symmetric solutions (\ref{1.6}), but we assume that both theories introduce small corrections to General Relativity in order to avoid violations of local gravity tests, as it has been pointed out in several works for $f(R)$ gravity (see Ref.~\cite{f(R)viable1}). We are interested to study  perturbations around a background solution in Brans-Dicke theory and  f(R) gravity (by means of its O'Halon description) via the Einstein frame, where the action (\ref{1.4}) can be transformed by performing a conformal transformation as it is well known. This also serves to compare the results within the Jordan and Einstein frames, where the possible non-physical equivalence between both frames is shown.

\section{Relation between $f(R)$ gravity and minimally coupled scalar-tensor theory via conformal transformations}

As it was pointed out in Ref.~\cite{Capozziello1}, the presence of a self-interacting scalar potential in O'Halon theory compared with BD theory, makes  the perturbations in the Newtonian limit of a spherically symmetric solution to not  be comparable, so that here we use the mathematically equivalent Einstein frame to study both theories and compare the results between the different frames. \\
In the Einstein frame,  the scalar field in the action is minimally coupled to the gravitational field, and the equations acquire a shape similar  to that of General Relativity. The relation between the Jordan and Einstein frames is given by the conformal transformation,
\be
g_{E\mu\nu}=\Pi^2g_{ \mu\nu},  \quad \text{where} \quad \Pi^2=\phi\ .
\label{2.1}
\ee
Then, the action (\ref{1.4})  in the Einstein frame becomes,
\be
S_{E}=\int d^4x\sqrt{-g_E}\left[R_E-\frac{2w+3}{2\phi^2}\nabla_{\mu}\phi\nabla^{\mu}\phi-\frac{V(\phi)}{\phi^2} +2\kappa^2\mathcal{L}_{Em}\right]\ ,
\label{2.2}
\ee
where the subscript $_E$ denotes that the variables are defined in the Einstein frame and the Lagrangian of matter is given by $\mathcal{L}_{Em}=\frac{1}{\phi^2}\mathcal{L}_m\left(\frac{1}{\phi^2}g_{E\mu\nu}\right)$. Note that O'Halon theory is described in the Einstein frame by the action (\ref{2.2}) with $w=0$, while Brans-Dicke theory is defined by $V(\phi)=0$ and $w\neq0$. In order to simplify the equations, we can redefine the scalar field  as $\phi=\e^{\varphi/\sqrt{3+2w}}$, and the action (\ref{2.2}) takes the form,
\be
S_{E}=\int d^4x\sqrt{-g_E}\left[R_E-\frac{1}{2}\nabla_{\mu}\varphi\nabla^{\mu}\varphi-U(\varphi) +2\kappa^2\mathcal{L}_{Em}\right]\ ,
\label{2.3}
\ee
here the scalar potential is defined as $U(\varphi)=\e^{\varphi/\sqrt{3+2w}}V(\phi(\varphi))$. The field equations can be   obtained by varying the action (\ref{2.3}) with respect to  $g_{E \mu\nu}$ and $\phi$,
\be
 R_{E\mu\nu}-\frac{1}{2}g_{E\mu\nu}R_E=\frac{1}{2}d_{\mu}\varphi d_{\nu}\varphi-\frac{1}{2}g_{E\mu\nu}\left[d_{\sigma}\varphi d^{\sigma}\varphi+U(\varphi)\right]+\kappa^2T^{(m)}_{E\mu\nu}\ ,
\label{2.4}
\ee
\be
\Box\varphi-\frac{dU(\varphi)}{d\varphi}=-2\kappa^2\frac{\delta(\mathcal{L}_{Em})}{\delta\varphi}\ ,
\label{2.5}
\ee
where the energy-momentum tensor is defined as $T_{E\mu\nu}^{(m)}=\frac{-2}{\sqrt{-g_E}}\frac{\delta\mathcal{L}_{E\mathrm{m}}}{\delta g_E^{\mu\nu}}$. By taking the trace of the first equation in (\ref{2.4}), the auxiliary equation is obtained,
\be
R_E=\frac{1}{2} d_{\sigma}\varphi d^{\sigma}\varphi+2U(\varphi)+\kappa^2T^{(m)}_E\ .
\label{2.6}
\ee
Hence, we will use the equations (\ref{2.4})-(\ref{2.6}) to study the perturbations created by the scalar field $\varphi$ in the Einstein frame around a background solution. Nevertheless, the spherically symmetric solution (\ref{1.6}) is transformed in the Einstein frame by the conformal factor (\ref{2.1}), and it  yields,
\be
ds_E^2=\Pi^2(r,t)ds^2=-\Pi^2A(r,t)dt^2+\Pi^2B(r,t)dr^2+\Pi^2r^2d\Omega\ .
\label{2.7}
\ee
By redefining  the coordinates, the metric can be rewritten in the more familiar form given in (\ref{1.6}),  taking $\rho^2=\Pi^2(r,t)r^2$, and $t'=T(t,r)$ to avoid cross terms, the metric (\ref{2.7}) is rewritten as,
\be
ds_E^2=-C(\rho,t')dt'^2+D(\rho,t')d\rho^2+\rho^2d\Omega\ .
\label{2.8}
\ee
Hence, we will  study the perturbations in the metric (\ref{2.8}) around a background solution, which will provide the range of validity of  Birkhoff's theorem in the Einstein frame in terms of the order of the perturbations, and for comparing the results  obtained for O'Halon and Brans-Dicke theory, as well as  the study the differences between the Einstein and Jordan frames.

\section{Spherically symmetric solutions and Birkhoff's theorem}

Let us now study the metric (\ref{2.8}) for the general scalar-tensor theory defined by the action (\ref{2.3}), and explore the range where such metric is static. It is well known that for Einstein's field equations, the  only solution in vacuum for a spherically symmetric metric is given by the Schwarzschild solution, or Schwarzschild-(A)dS solution if a cosmological constant is included in the field equations. This result, called Birkhoff's theorem, was proved independently by G. D. Birkhoff \cite{Birkhoff} and J. T Jebsen \cite{Jebsen}, and it states the following,\\

\textbf{Birkhoff's theorem}: \textit{a spherically symmetric solution of the vacuum Einstein equations is always static in a region where the time coordinate remains time-like and spatial coordinates stay space-like.} \\

Here we want to explore the range of validity of this theorem for $f(R)$ gravity and Brans-Dicke theory trough out their description in the Einstein frame given by the action (\ref{2.3}) and the field equations (\ref{2.4})-(\ref{2.6}). It was pointed out in Ref.~\cite{Capozziello3} that in f(R) gravity, the Birkhoff's theorem is valid up to fourth order in perturbations in the Newtonian limit of the theory, what implies the weak field regime. Nevertheless, for other regimes, where the scalar curvature is sufficiently large, such limit would not be valid, and the solution can change. Here, we consider a more general context, where a background solution for the metric (\ref{2.8}) and the scalar field is considered, and perturbations around the zero-order solution are studied. Then, the metric and scalar field can be written as,
\be
g_{E\mu\nu}=g_{E\mu\nu}^{(0)}+g_{E\mu\nu}^{(1)}\ , \quad \varphi=\varphi^{(0)}+\varphi^{(1)}\ .
\label{3.1}
\ee
Here $g_{E\mu\nu}^{(0)}$ refers to the zero-order solution, while the perturbations are represented by $g_{E\mu\nu}^{(1)}$. By taking the metric (\ref{2.8}), the components of the metric can be written as
\be
\left\{\begin{array}{ll}
g_{Et't'}=-C(\rho,t')\simeq -C^{(0)}(\rho,t')-C^{(1)}(\rho,t') \\
g_{E\rho\rho}=D(\rho,t')\simeq D^{(0)}(\rho,t')+D^{(1)}(\rho,t') \\
g_{E\theta\theta}=\rho^2 \\
g_{E\psi\psi}=\rho^2 \sin^2\theta
\end{array}\right.
\label{3.2}
\ee
while the inverse metric, defined as usually by $g_{E}^{\alpha\sigma}g_{E\sigma\lambda}=\delta^{\alpha}_{\lambda}$,  yields,
\be
\left\{\begin{array}{ll}
g^{t't'}_E=-\frac{1}{C(\rho,t')}\simeq -\frac{1}{C^{(0)}(\rho,t')}+\frac{C^{(1)}(\rho,t')}{(C^{(0)}(\rho,t'))^2} \\
g^{\rho\rho}_E=\frac{1}{D(\rho,t')}\simeq \frac{1}{D^{(0)}(\rho,t')}-\frac{D^{(1)}(\rho,t')}{(D^{(0)}(\rho,t'))^2}  \\
g^{\theta\theta}_E=\frac{1}{\rho^2} \\
g^{\psi\psi}_E=\frac{1}{\rho^2 \sin^2\theta}
\end{array}\right.
\label{3.3}
\ee
The Ricci tensor can be written in its standard form as,
\be
R_E=g_E^{\mu\nu}R_{E\mu\nu}= g_E^{\mu\nu}\left(\Gamma^{\alpha}_{\mu\nu,\alpha}-\Gamma^{\alpha}_{\mu\alpha,\nu}+\Gamma^{\beta}_{\beta\alpha}\Gamma^{\alpha}_{\mu\nu}-\Gamma^{\alpha}_{\beta\mu}\Gamma^{\beta}_{\nu\alpha}\right)\ ,
\label{3.4}
\ee
where the Christoffel symbols are defined by,
\be
\Gamma^{\alpha}_{\mu\nu}=\frac{1}{2}g_E^{\alpha\sigma}\left(g_{E\mu\sigma,\nu}+g_{E\nu\sigma,\mu}-g_{E\mu\nu,\sigma}\right)\ .
\label{3.5}
\ee
By the perturbed metric defined in (\ref{3.2}) and (\ref{3.3}), we can split the Christoffel symbols into two orders of perturbations, $\Gamma^{\alpha}_{\mu\nu}={\Gamma^{(0)}}^{\alpha}_{\mu\nu}+{\Gamma^{(1)}}^{\alpha}_{\mu\nu}$, as well as the Ricci scalar $R_E=R^{(0)}_E+R^{(1)}_E$. The perturbations also act on the scalar potential $U(\varphi)$, such that the potential can be expanded around a background solution for the scalar field $\varphi_0$,
\be
U(\varphi)=\sum_n \frac{U^{n(0)}}{n!}\left(\varphi-\varphi^{(0)}\right)^n\ ,
\label{3.6}
\ee
By inserting the above expressions in the field equations (\ref{2.4})-(\ref{2.6}), we can split the equations into the different orders of perturbations. We are interested in vacuum solutions,  $T_{E\mu\nu}^{(m)}=0$, and we assume that the background solution is given by a constant scalar field $\varphi^{(0)}(\rho,t')=\varphi_0$. The reason for this is that it is assumed that at zero-order the results for General Relativity has to be recovered, and the additional terms in the action are involved at higher orders only. Then, the equations (\ref{2.4}) and (\ref{2.5}) at zero-order are given by,
\be
R_{E\mu\nu}^{(0)}-\frac{1}{2}g_{E\mu\nu}^{(0)}R_E^{(0)}+g_{E\mu\nu}^{(0)} \Lambda=0
\label{3.7}
\ee
\be
\frac{dU(\varphi^{(0)})}{d\varphi}=0\ ,
\label{3.8}
\ee
where the cosmological constant is defined as $\Lambda=\frac{1}{2}U_0$. The equations (\ref{3.7}) are exactly the same as the Einstein field equations with a cosmological constant, and the solution is the well known Schwarzschild-(A)dS metric, that gives the zero-order solution,
\be
C^{(0)}(\rho)=[D^{(0)}(\rho)]^{-1}=1-\frac{2\mu}{\rho}-\frac{\Lambda}{3} \rho^2\ ,
\label{3.8a}
\ee
where $\mu$ is an integration constant. Then, at zero-order we have a static solution which satisfies Birkhoff's theorem as expected. At first linear order, the equations (\ref{2.4})-(\ref{2.6}) are,
\be
R^{(1)}_{E\mu\nu}-\frac{1}{2}\left(g^{(0)}_{E\mu\nu}R^{(1)}+g^{(1)}_{E\mu\nu}R^{(0)}\right)=\frac{1}{2}U_0'g^{(0)}_{E\mu\nu}\varphi^{(1)}-\frac{1}{2}U_0g^{(1)}_{E\mu\nu}\ ,
\label{3.9}
\ee
\be
\Box\varphi^{(1)}=U_0'' \varphi^{(1)}\ ,
\label{3.10}
\ee
\be
R^{(1)}=2U_0'\varphi^{(1)}\ .
\label{3.11}
\ee
Introducing the results obtained at zero order (\ref{3.8}), where we have $U_0'=0$, and $R^{(0)}=2U_0$, the equations (\ref{3.9})-(\ref{3.11}) turn out much simpler, where the field equations for the metric (\ref{3.9}) yield,
\be
R^{(1)}_{E\mu\nu}-\frac{1}{2}U_0g^{(1)}_{E\mu\nu}=0\ .
\label{3.12}
\ee
Note that  the scalar field decouples from the metric at linear order, which is not expected at higher orders. The expression (\ref{3.12}) is a linear system of differential equations, where the coefficients are given in terms of the zero order  (\ref{3.8a}), and whose solution exhibits a very complex expression (for a  complete analysis on the perturbations of a black hole in $f(R)$ gravity, see Ref.~\cite{arXiv:1103.0343}). Nevertheless, we can study the system in order to obtain the enough information about the form of the metric components at first linear order and their staticity. From the $rt$-equation,
\be
R_{tr}^{(1)}=\frac{1}{r}\frac{\dot{D}^{(1)}}{D^{(0)}}=0\, \quad \rightarrow \quad  g^{(1)}_{E\rho\rho}=D^{(1)}(\rho,t')=D^{(1)}(\rho)\ .
\label{3.12a}
\ee
Hence, the $rr$-component of the metric is time independent. By deriving the $\theta\theta$-equation respect the time, it gives,
\be
\frac{dR^{(1)}_{\theta\theta}}{dt}=-\frac{r}{2D^{(0)}}\frac{d^2}{drdt}\left(\frac{C^{(1)}}{C^{(0)}}\right)=0\ , \quad \rightarrow C^{(1)}(\rho,t')=C^{(0)}(\rho)\left(\alpha(t')+\chi(\rho)\right)\ ,
\label{3.12b}
\ee
It is straightforward to show by the $rr$- and  $tt$-equations that $\alpha(t')$ remains an undetermined function of the time $t'$, so that it can be taken to be constant $\alpha(t')=\alpha$, while the functions $D^{(1)}(\rho)$ and $\chi(\rho)$ are solutions of the system of differential equations  (\ref{3.12}).  In general, one can not obtain a general solution for $D^{(1)}(\rho)$ and $\chi(\rho)$, but numerical methods are required. However, the solutions (\ref{3.12a}-\ref{3.12b}) provide the enough information to analyze the metric components. As they  are time-independent,  Birkhoff's theorem is satisfied at this order by the solution, and the metric remains static in vacuum for a general scalar-tensor theory described by the action (\ref{2.3}) in the Einstein frame.  \\
At this point we have to remark the differences between O'Halon and Brans-Dicke theory described both in the Einstein frame by the action (\ref{2.3}). It is straightforward to show that for BD theory, where the scalar potential $U(\varphi)=0$, the solutions become quite different compared with O'Halon theory. At zero-order, the field equations (\ref{3.7}) are different for each theory, while in BD the cosmological constant vanishes and the zero-order solution (\ref{3.8a}) turns out the classical Schwarzschild solution,  for O'Halon theory the metric  is given by Schwarzschild-(A)dS solution at zero order. At first linear order, the equations (\ref{3.9}) reduces to $R^{(1)}_{E\mu\nu}=0$ in BD theory, whose solution is also static, but naturally it gives a different metric compared with O'Halon theory, as it was expected. Note that up to this point the only term used to compare both theories is the scalar potential, while the parameter $w$ does not play any role in the Einstein frame, but on the inverse conformal transformation to turn back to the Jordan frame. Then, this provides a new method to compare both theories in perturbations using the Einstein frame representation. In this case, we have found that Birkhoff's theorem is satisfied in both approaches at least  to first linear order, although the solutions for the metric are different for each approach.    \\

Let us now transform the metric components obtained in the Einstein frame (\ref{3.12a}-\ref{3.12b}) to the Jordan one, and check how the metric is affected. In order to obtain the conformal transformation (\ref{2.1}), we have to solve the scalar field equation at first linear order (\ref{3.10}), which can be extended,
\be
C^{(0)}(\rho)\frac{d^2\varphi^{(1)}}{d\rho^2}-\left(C'^{(0)}+\frac{2}{r}C^{(0)}\right)\frac{d\varphi^{(1)}}{d\rho}-\frac{1}{C^{(0)}(\rho)}\frac{d^2\varphi^{(1)}}{dt^2}-U_0''\varphi^{(1)}=0\ .
\label{3.14a}
\ee
This equation clearly has a solution where we can split up the time and radial parts of the scalar field,
\be
\varphi^{(1)}(\rho,t')=\varphi^{(1)}_{\rho}\varphi^{(1)}_{t'}\ .
\label{3.14}
\ee
And the equation (\ref{3.14a}) takes the form,
\be
C^{(0)}\varphi''^{(1)}_{\rho}-\left(C'^{(0)}+\frac{2}{r}C^{(0)}\right)\varphi'^{(1)}_{\rho}-U_0''\varphi^{(1)}_{\rho}-\frac{\varphi^{(1)}_{\rho}}{C^{(0)}}\frac{\ddot{\varphi}^{(1)}_{t'}}{\varphi^{(1)}_{t'}}=0\ .
\label{3.14b}
\ee
Here, the primes denote derivatives with respect to the radial coordinate while the dots denote with respect to the time coordinate. Hence, the corresponding time solution of the equation has to hold,
\be
\frac{\ddot{\varphi}^{(1)}_{t'}}{\varphi^{(1)}_{t'}}=k\, \quad \rightarrow \quad \varphi^{(1)}_{t'}=C_1\e^{\sqrt{k} t'}+ C_2\e^{-\sqrt{k} t'}\ ,
\label{3.14c}
\ee
where $C_{1,2}$ and $k$ are constants. Note that the sign of $k$ may play an important role on the stability of the background solution in the Jordan frame, as an exponential function is obtained for a positive value of $k$, while an oscillator is obtained for a negative $k$. The radial part of the solution $\varphi^{(1)}_{\rho}$ is obtained  by solving the differential equation,
\be
 C^{(0)}\varphi''^{(1)}(\rho)+\left[C'^{(0)}+\frac{2}{r}C^{(0)}\right]\varphi'^{(1)}(\rho)-\left(\frac{k}{C^{(0)}}+U_0''\right)\varphi^{(1)}(\rho)=0\ .
\label{3.15}
\ee
Hence, the conformal transformation yields,
\be
\Pi^2=\phi=\phi^{(0)}+\phi^{(1)}=\e^{\varphi/\sqrt{3+2w}}\simeq \phi^{(0)} \left(1+\frac{1}{\sqrt{3+2w}}\varphi^{(1)}\right)\ ,
\label{3.16}
\ee
where $\phi^{(0)}=\e^{\varphi^{(0)}/\sqrt{3+2w}}$. Recall that O'Halon theory is defined by $w=0$, so the inverse of the conformal transformation (\ref{3.16}) in BD and O'Halon theories differs only in a numerical factor, and the same differences between both theories that were found in the Einstein frame, remain in the Jordan one. However, the important result here refers to the differences between the Einstein and Jordan frames. By the transformation (\ref{3.16}), it is clear  that the perturbations on the metric in the Jordan frame will not be static at first linear order, and the Birkhoff's theorem will not be valid in this frame, which differs completely from the Einstein one. Note also that, in general the result given in (\ref{3.14c}), suggests that the zero-order solution will be unstable in the Jordan frame due to the perturbations introduced by the freedom of the scalar field. This fact suggests the non-physical equivalence between both frames, as it has already pointed out in Ref.~\cite{JordanVsEinstein} by the analysis of both frames in a cosmological context. The transformed metric in the Jordan frame is given by,
\be
ds^2=-\frac{C(\rho)}{\phi(\rho,t')}dt'^2+\frac{D(\rho)}{\phi(\rho,t')}d\rho^2+\frac{\rho^2}{\phi(\rho,t')}d\Omega\ .
\label{3.17}
\ee
At zero order, where $\phi^{(0)}$ in (\ref{3.16}) is a constant, the result in the Jordan frame gives also a Schwarzschild-(A)dS metric, and the Birkhoff's theorem is satisfied. Nevertheless, at the first linear order the metric is clearly not static, which contradicts the result obtained in the Einstein frame, and points to the different physical meaning of both frames. By redefining the radial coordinate $r^2=\phi(\rho,t')\rho^2$ in (\ref{3.17}), and  performing some variable transformations, the metric (\ref{3.17}) can be written in the more common form given in (\ref{1.6}), where the temporal and radial components of the metric will depend on time, introducing a strong instability on the background solution.

\section{Discussions}

In summary, we have studied the range of validity of Birkhoff's theorem in $f(R)$ gravity through its O'Halon representation, and in Brans-Dicke theory, both expressed in the Einstein frame. The comparison with the results in the Jordan frame has been also analyzed. Here we have presented a method to compare the two theories in the Einstein frame instead the Jordan one, where both theories have to be studied separately in the perturbations in order to obtain consistent results, as  pointed out in Ref.~\cite{Capozziello3}. We have seen that assuming a background solution described by a constant scalar field (as a limit to General Relativity),  the zero-order solution gives naturally a static metric, the (A)dS-Schwarzschild solution. However, also the first linear order provides a metric that is time-independent in the Einstein frame, so that the Birkhoff's theorem is satisfied. Nevertheless, the fact that Brans-Dicke theory has a null scalar potential makes the solution different compared with  O'Halon theory, although the theorem is also satisfied, at least for first linear order in perturbations. When non-linear terms are included, it is expected that the metric will not remain static in both theories. \\

In parallel, we found that the result obtained in the Einstein frame about the range of validity of the Birkhoff's theorem  is affected when a conformal transformation is applied, and the action is transformed to the Jordan frame, where the metric is not static at first linear order in perturbations. Hence, this can mean that  the Jordan and the Einstein frames are not physically equivalents, at least for a perturbative analysis. While in the Einstein frame, Birkhoff's theorem is valid at first linear order, where the scalar field remains decoupled, the conformal transformation makes the metric depends on time when it is transformed to the Jordan frame, where the coupling with the scalar field is recovered, violating the Birkhoff's theorem. In addition, the presence of a time-dependence in the metric can affect  the stability of the background solution, which was found to be stable in the Einstein frame. Then, as pointed out in Ref.~\cite{JordanVsEinstein}, both frames may be mathematically but non-physically equivalent.

\section*{Acknowledgments}
The Authors wish to thank S.D. Odintsov for common discussions and suggestions on the topics. Part of DSG's work was carried out during a stay  at the Universit\`a di Napoli ``Federico II'', and would like to thank the Department of Physics for kind hospitality. DSG acknowledges an FPI fellowship from MICINN (Spain), project FIS2006-02842.

\end{document}